\documentclass[sigconf]{acmart}

\usepackage{pifont}
\usepackage{enumitem}
\usepackage{makecell}
\usepackage{multirow}
\usepackage[ruled,linesnumbered]{algorithm2e}
\usepackage{algorithmic}
\usepackage{subcaption}
\usepackage{pifont}
\usepackage{bbding}
\usepackage{colortbl}
\usepackage{soul}
\usepackage{booktabs}
\usepackage{xcolor}
\usepackage{tikz}
\usepackage{environ}
\definecolor{grey}{rgb}{0.9,0.9,0.9}
\newenvironment{answerbox}[1]{
  \setlength{\parindent}{0pt}
  \setlength{\rightskip}{0pt}
  \begin{tikzpicture}[baseline=(mybox.base)]
    \node[
        draw=black,
        fill=grey,
        rounded corners=5pt, 
        inner sep=.5em,      
        text width=\linewidth-1em, 
        align=justify
    ] (mybox) {#1}; 
  \end{tikzpicture}
}

\definecolor{customgray}{rgb}{0.651, 0.651, 0.651}
\definecolor{plausible_origin}{RGB}{255,131,52}
\definecolor{correct_green}{RGB}{58,140,41}
\newcommand{\correct}{\textcolor{correct_green}{\ding{51}}}
\usepackage{tikz}
\def\plausible{\textcolor{plausible_origin}{\ding{51}}{\small\textcolor{plausible_origin}{\kern-0.7em\ding{55}}}}
\def\correct{\textcolor{correct_green}{\CheckmarkBold}}

\AtBeginDocument{%
  }

\copyrightyear{2026}
\acmYear{2026}
\setcopyright{cc}
\setcctype{by}
\acmConference[ICSE '26]{2026 IEEE/ACM 48th International Conference on Software Engineering}{April 12--18, 2026}{Rio de Janeiro, Brazil}
\acmBooktitle{2026 IEEE/ACM 48th International Conference on Software Engineering (ICSE '26), April 12--18, 2026, Rio de Janeiro, Brazil}
\acmPrice{}
\acmDOI{10.1145/3744916.3773218}
\acmISBN{979-8-4007-2025-3/26/04}

\begin{document}

\newcommand{\sysname}{\textsc{LoopRepair}\xspace}
\title{Well Begun is Half Done: Location-Aware and Trace-Guided Iterative Automated Vulnerability Repair}
\thanks{\\
*Xiaobing Sun and Sicong Cao are the corresponding authors.\\
}





\author{Zhenlei Ye}
\affiliation{%
  \institution{Yangzhou University}
  \city{Yangzhou}
  \country{China}
}
\email{DX120240105@stu.yzu.edu.cn}

\author{Xiaobing Sun}
\authornotemark[1]
\affiliation{%
  \institution{Yangzhou University}
  \city{Yangzhou}
  \country{China}}
\email{xbsun@yzu.edu.cn}

\author{Sicong Cao}
\authornotemark[1]
\affiliation{%
  \institution{Nanjing University of Posts and Telecommunications}
  \city{Nanjing}
  \country{China}
}
\email{sicong.cao@njupt.edu.cn}

\author{Lili Bo}
\affiliation{%
 \institution{Yangzhou University}
  \city{Yangzhou}
  \country{China}
}
\email{lilibo@yzu.edu.cn}

\author{Bin Li}
\affiliation{%
 \institution{Yangzhou University}
  \city{Yangzhou}
  \country{China}
}
\email{lb@yzu.edu.cn}


\begin{abstract}
  The advances of large language models (LLMs) have paved the way for automated software vulnerability repair approaches, which iteratively refine the patch until it becomes plausible. Nevertheless, existing LLM-based vulnerability repair approaches face notable limitations: 1) they ignore the concern of locations that need to be patched and focus solely on the repair content. 2) they lack quality assessment for generated candidate patches in the iterative process.
  
  To tackle the two limitations, we propose \sysname, an LLM-based approach that provides information about where should be patched first. Furthermore, \sysname improves the iterative repair strategy by assessing the quality of test-failing patches and selecting the best patch for the next iteration. We introduce two dimensions to assess the quality of patches: whether they introduce new vulnerabilities and the taint statement coverage. We evaluated \sysname on a real-world C/C++ vulnerability repair dataset VulnLoc+, which contains 40 vulnerabilities and their Proofs-of-Vulnerability. The experimental results demonstrate that \sysname exhibits substantial improvements compared with the Neural Machine Translation-based, Program Analysis-based, and LLM-based state-of-the-art vulnerability repair approaches. Specifically, \sysname is able to generate 27 plausible patches, which is comparable to or even 8 to 22 more plausible patches than the baselines. In terms of correct patch generation, \sysname repairs 8 to 13 additional vulnerabilities compared with existing approaches.
\end{abstract}




\maketitle

\section{Introduction}
With the explosion of open-source software (OSS), disclosed OSS vulnerabilities have tremendously increased~\cite{ye2025kg4va,sun2025hgtjit,DBLP:conf/uss/Gao0C0WLLX25,DBLP:conf/kbse/Cao000B00L024}. The number of disclosed vulnerabilities reaches as high as 36,109 in 2024, as reported by the National Vulnerability Database (NVD)~\cite{NVD}. However, a recent study indicates that fixing a vulnerability takes a median of 270 days~\cite{fixingspeed}. 
These alarming statistics indicate the critical need for software vulnerability repair to minimize the time window of opportunity for potential exploits \cite{Xiao,exp,DBLP:conf/kbse/0001GCB0024}.
To tackle this issue, conventional approaches primarily employ static/dynamic analysis techniques to generate security patches~\cite{MemFix,DBLP:journals/usenix-login/HuangLTJ20,ExtractFix,SAVER,EffFix,DBLP:conf/pldi/ShariffdeenNGR21}, making them less suitable for addressing all types of vulnerabilities. Recently, researchers have increasingly focused their attention on learning-based Automated Vulnerability Repair (AVR) tasks, which can generate patches for all types of vulnerabilities.

Some researchers focus on Neural Machine Translation (NMT)-based AVR, which aims to learn repair patterns from a large corpus of <vulnerable function, patch function> pairs~\cite{VRepair,VulRepair,VQM,VulMaster}. They treat the AVR task as a translation task, which trains a sequence-to-sequence model that inputs the vulnerable function and outputs the patch function. Specifically, they employ \textit{cloze-style} repair strategy~\cite{AlphaRepair}, which masks the \textit{perfect vulnerability localization}~\cite{VulMaster} and uses a well-trained model to predict the token sequence in the masked locations. However, NMT-based approaches fail to solve multi-hunk vulnerabilities, which are defined as vulnerabilities that require edits at multiple, non-contiguous hunks in a program. It is because the locations that need to be patched are not the same as the locations that are vulnerable.
Some Large Language Model (LLM)-based approaches have been proposed to solve this problem~\cite{AlphaRepair,ChatRepair,ThinkRepair,ITER}. Their main idea is to transform a single repair attempt into multiple repair attempts, addressing a part of the vulnerability each time and iteratively repairing the generated patches until the multi-hunk vulnerability is successfully fixed~\cite{ITER}. Due to the conversational abilities of LLMs, many studies~\cite{ChatRepair,ThinkRepair,ITER} have adopted an iterative repair strategy that continuously utilizes LLMs to refine failing patches until they become plausible. However, the LLM-based iterative strategy still faces two major challenges:

\ding{172} \textbf{Ignoring the concern of locations that need to be patched.} 
Within each iteration of the iterative repair process, the LLM generates the candidate patches based on the vulnerable function and vulnerable hunk locations provided by the localization tools.
However, they overlook an important fact: \textit{The location of the vulnerable hunks are not always aligned with the locations that need to be patched, meaning that fixing a vulnerability in one location may require modifications in multiple places. Therefore, it is important for LLMs to pay attention to the locations that need to be repaired before generating patches}. For example, for a Use After Free (CWE-416~\cite{CWE-416}) vulnerability, it is exploited because a pointer is called after it has been freed. In order to repair this vulnerability, we need to understand which pointer triggered the vulnerability and locate all instances of that pointer, setting it to \texttt{NULL} afterward. If we cannot identify which hunks need to be patched before generating the patches, the vulnerability is hard to be successfully repaired and further decreases the performance of AVR.

\ding{173} \textbf{Lack of assessment for the quality of generated patches.} 
On the one hand, existing LLM-based iterative strategies do not verify whether generated patches introduce new vulnerabilities, which means that if the repair tool introduces new vulnerabilities without fixing the original ones, the generated patches could contain multiple vulnerabilities. This situation can severely impact the effectiveness of the iterative vulnerability repair process.
On the other hand, existing iterative approaches directly refine the generated patches rather than assess the quality of generated patches.
For example, if both the patch $\textit{n}$ (generated in $\textit{n}^{th}$ iteration) and the patch $(\textit{n+1})$ (generated based on patch $\textit{n}$) cannot pass Proof-of-Vulnerability (PoV) verification, existing strategies will directly employ next iteration based on patch $(\textit{n+1})$. However, there is no evidence to show that patch $(\textit{n+1})$ is better than patch $\textit{n}$.
This indicates that existing approaches lack the necessary quality assessment and iterative repair based on patches of unknown quality can impact the performance of the AVR.

To tackle the above challenges, we propose \sysname, a novel automated vulnerability repair approach that aims to improve the iterative strategy with \underline{lo}cation-aware patch generati\underline{o}n and taint trace-guided \underline{p}atch selection. \sysname consists of two major components. First, to provide information about where needs to be patched, it treats the patch hunk location prediction task as a translation task and employs an LLM to generate the line number sequence which contains the information about where needs to be patched. Second, to assess the quality of generated patches, it employs taint trace, dynamic runtime information that captures the executed taint statements and the data flow from taint source to the taint sink. To assess whether the generated patches introduce new vulnerabilities, it compares the taint sink between the generated patch and the original program. Under the same vulnerability-triggering input, if they share the same taint sink and the same CWE-ID, this indicates that there is no vulnerability introduction. To assess the quality of each patch, it calculates the taint statement coverage for each patch, and the top-ranked patch is then selected for the next iteration. Our key insight is that, given the same taint source, the highest taint statement coverage means that the patch has the greatest impact on vulnerability exploitation compared with other patches. Therefore, continuously repairing this patch is more likely to effectively disrupt the exploitation path and successfully fix the vulnerability than other patches.

We evaluate the performance of \sysname on the VulnLoc+~\cite{VulnLoc,CrashRepair} dataset, which consists of 40 vulnerabilities with their PoVs collected from 10 open-source software projects. \sysname is able to generate 27 plausible patches, which is comparable to the baseline~\cite{CrashRepair} or even 8 to 22 more plausible patches than other baselines~\cite{VRepair,VulRepair,VulMaster,VQM,ChatRepair,ThinkRepair,ITER,VulnFix}. In terms of correct patch generation, \sysname achieves state-of-the-art performance, repairing 8 to 13 additional vulnerabilities compared with existing approaches. 
In summary, our contributions are as follows.
\begin{itemize}[leftmargin=1em]
    \item We propose a novel approach \sysname that enhances AVR via location-aware patch generation and taint trace-guided patch selection.
    
    \item We design a taint trace-guided iterative repair strategy that ranks candidate patches based on whether they introduce new vulnerabilities and their taint statement coverage, selecting the best patch for iterative repair.
    
    \item We conduct experiments on the VulnLoc+ dataset, and the experimental results show that \sysname outperforms all baselines. 
    The source code, dataset, and generated patches in our experiments are available at \url{https://github.com/Fino2020/LoopRepair}.
\end{itemize}

\section{Basics and Motivation}
\subsection{Definitions}\label{TaskDefinition}

\underline{\textbf{Task Definition}}. In this paper, we focus on function-level Automated Vulnerability Repair (AVR), which takes a sequential vulnerable function $F_{v}$ and the vulnerable statement $S_{v}$ as inputs, and outputs the patch function $F_{p}$. Following previous work~\cite{CrashRepair,ExtractFix}, we assume that the vulnerability has already been detected and that 1-2 Proofs-of-Vulnerability (PoVs) are available.

\noindent \underline{\textbf{Multi-Hunk Vulnerability}}. We refer to one or several consecutive statements as a single hunk. A multi-hunk vulnerability is a vulnerability that requires edits at multiple, non-contiguous hunks in a program~\cite{ITER,Indivisible}. 

\subsection{Motivation}\label{motivation}

\begin{figure}[!t]
    \centering 
    \includegraphics[width=\linewidth]{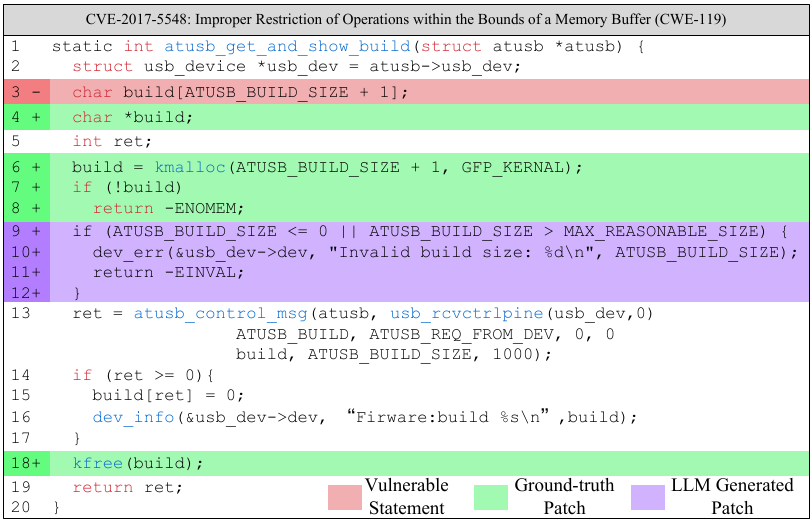}
  \caption{A multi-hunk vulnerability example (CVE-2017-5548~\cite{CVE-2017-5548})
  , which belongs to CWE-119. The vulnerable hunk is marked in red, the patch hunks are marked in green, and the patch generated by LLM is marked in purple.
  }
  \label{fig:background}
\end{figure}

Figure~\ref{fig:background} shows a typical \textit{Improper Restriction of Operation within the Bounds of a Memory Buffer} vulnerability CVE-2017-5548~\cite{CVE-2017-5548} in \textit{Linux Kernel}~\cite{LinuxKernel}. It is a multi-hunk vulnerability, which is caused by the incompatibility between stack memory allocation (i.e., the stack buffer \texttt{build}) and Direct Memory Access (DMA) operations. The vulnerable hunk is marked in red, while its ground-truth patch hunks are marked in green. In order to successfully repair this vulnerability, the ground-truth patch replaces the stack buffer (\texttt{build} at \underline{Line 3}) with a dynamically allocated heap buffer (\texttt{kmalloc} at \underline{Line 6}). To convert the stack buffer \texttt{build} to a heap buffer, all following \texttt{build}-related operations need to be modified. This means that to automatically repair this vulnerability, the repair tools need to remove one vulnerable hunk (i.e., \underline{Line 3}) and add three patch hunks (i.e., \{\underline{Line 4}, \underline{Lines 6-8}, \underline{Line 18}\}). 

\noindent\textbf{Observation 1. The vulnerable hunk locations do not align with the patch hunk locations, highlighting the importance of providing the patch hunk locations before generating patches.}\label{motivation1} Let us review how existing approaches fail to solve multi-hunk vulnerabilities. (1) For Neural Machine Translation (NMT)-based approaches~\cite{VRepair,VulRepair,VulMaster,VQM,Angelix}, they mask vulnerable hunks and utilize a well-trained model to generate a token sequence within the masked locations. For example, to generate patches for the vulnerable function in Figure~\ref{fig:background}, the hunk at \underline{Line 3} will be masked as <\textit{Vul-Start}>$ char\;build[ATUSB\_BUILD\_SIZE\;+\;1];$<\textit{Vul-End}>. The repair model is designed to generate the token sequence between the start token <\textit{Vul-Start}> and the end token  <\textit{Vul-End}>. 
However, the following \texttt{build}-related hunks will not be modified. 
It indicates that NMT-based approaches can hardly repair multi-hunk vulnerabilities.
(2) For iterative approaches based on the Large Language Model (LLM), they employ iterative strategies that highlight that each repair attempt can fix a portion of the vulnerability and continuous iterative repair can address multi-hunk vulnerabilities~\cite{ITER}. Within each iteration, the LLM generates the candidate patches based on the vulnerable function and vulnerable hunk locations provided by the localization tools.
However, the locations of the vulnerable hunks and the patch hunks are not always aligned. Therefore,  \textit{it is important for LLMs to pay attention to the locations that need to be repaired before generating patches}. For example, for the vulnerable function shown in Figure~\ref{fig:background}, if we do not provide information about patch hunk locations, LLM will consider that the vulnerability is caused by not checking the \texttt{ATUSB\_BUILD\_SIZE} variable, leading to an array out-of-bounds issue. The LLM will then simply add an \texttt{if}-condition (marked in purple), but this patch does not successfully repair the vulnerability.

\begin{table}
    \centering
    \caption{The statistic of single-hunk vulnerabilities and multi-hunk vulnerabilities in VulRD and MegaVul datasets.}
    \resizebox{0.47\textwidth}{!}{
    \begin{tabular}{lcccr}
    \toprule
    Datasets & Total & \makecell{Single-Hunk\\ Vulnerability} & \makecell{Multi-Hunk\\Vulnerability} & \makecell{Percentage of \\ Multi-Hunk} \\  
    \midrule
    VulRD~\cite{Big-Vul,CVEFixes}  & 4,536   & 1,800   & 2,736    & 60.32\% \\
    MegaVul~\cite{MegaVul}   & 10,555    & 4,853    & 5,702     & 54.02\% \\
    \bottomrule
    \end{tabular}}
    \label{Multi-Hunk}
\end{table}

Actually, the above case is common in practice. We conduct a preliminary study to quantitatively analyze the multi-hunk vulnerabilities on two vulnerability-fixing datasets: VulRD~\cite{Big-Vul,CVEFixes} and MegaVul~\cite{MegaVul}. VulRD is widely used in previous work~\cite{VRepair,VulRepair,VulMaster} and comprises 4,536 C/C++ vulnerability-patch pairs by merging two existing datasets: Big-Vul~\cite{Big-Vul} and CVEFixes~\cite{CVEFixes}. MegaVul~\cite{MegaVul} comprises 10,015 C/C++ vulnerability-patch pairs. The results are shown in Table~\ref{Multi-Hunk}. It can be observed that the percentage of samples with multi-hunk vulnerabilities is 60.32\% and 54.02\%, respectively. This means that the multi-hunk vulnerabilities are very common in real-world scenarios, which seriously challenges the existing AVR approaches.

This inspires us to focus on the patch hunk locations that need to be patched and provide them to LLMs for better performance.
The most straightforward solution is to employ a model that inputs the vulnerable function along with the vulnerable hunk locations provided by the localization tool. This model then outputs the locations that should be patched.
The core insight behind this is that \textit{in order to successfully repair a vulnerability, we not only need to predict the content of the hunks, but more importantly, we need to know which hunks need to be repaired beforehand}. In this way, we can address the misalignment between vulnerable hunk locations and patch hunk locations and help LLMs focus on the patch hunk locations before generating patches.

\noindent\textbf{Observation 2. Iterative repair based on generated patches with unknown quality, highlighting the necessary patch quality assessment.}\label{motivation2}
Let us review how existing LLM-based iterative repair strategies (e.g., ITER~\cite{ITER}, ChatRepair~\cite{ChatRepair}, and ThinkRepair~\cite{ThinkRepair}) work. Figure~\ref{fig:motivation2} illustrates the workflow of iterative repair strategies, which takes a vulnerable function as input and iteratively refines the generated candidate patches. 
For each iteration, the candidate patch will be executed one repair attempt and then verified. The generated patches can be classified into three types: \ding{182} patch that fails to compilation, which is marked with a red circle. \ding{183} patch that can pass compilation but fails to pass Proofs-of-Vulnerability, which is marked with a blue triangle. \ding{184} patch that can pass compilation and PoV (i.e., plausible patch), which is marked with a green star.
The failing patches will be treated as input for the next iteration until they become plausible and correct. Existing iterative repair strategies operate under the concept that \textit{even failing patches can yield valuable insights}~\cite{ITER}. For example, the iterative process of Patch \texttt{0-z} → Patch \texttt{1-z} → Patch \texttt{2-z} → \texttt{$\cdots$} → Patch \texttt{n-y} in Figure~\ref{fig:motivation2} demonstrates that iterative repair of PoV-failing/compilation-failing patches could ultimately yield plausible patches. Although iterative repair strategies have been proved effective~\cite{ITER,ThinkRepair,ChatRepair}, they still have two aspects that need to be improved:

\begin{figure}[!t]
    \centering
    \includegraphics[width=0.47\textwidth]{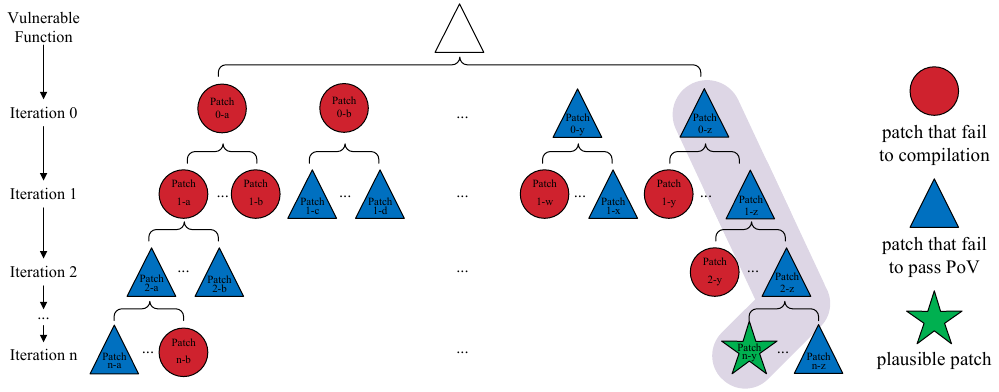}
    \caption{Iterative approaches (ITER~\cite{ITER}, ThinkRepair~\cite{ThinkRepair}, ChatRepair~\cite{ChatRepair}) execute repair operation on failing patches.}
    \label{fig:motivation2}
\end{figure}

\underline{\textbf{Avoiding new vulnerability introduction.}} 
Existing LLM-based iterative repair strategies cannot ensure that PoV-failing patches avoid introducing new vulnerabilities. Let us assume a scenario where a patch fails to pass PoV verification while introducing a new vulnerability. Under a scenario where most of the generated patches (more than 98\% in our experiments) fail to pass PoV verification, this patch cannot be identified among all the failing patches. Furthermore, due to the fact that there are only one or two available PoV(s) for each vulnerability in real-world scenarios, we cannot check whether the introduced new vulnerability is repaired in the following iterations. This indicates that failing patches that introduce new vulnerabilities not only increase repair complexity, but also degrade the reliability of subsequently generated patches. 
This motivates us to investigate whether the generated patches introduce new vulnerabilities during the iterative process and filter out suspicious patches. To achieve this, we designed a two-step criterion. First, if the failed patch shares the same CWE-ID as the original program, it can be confirmed that no other types of vulnerabilities have been introduced. Second, under the condition of using the same PoV, if the taint sink of the failed patch is the same as that of the original program, the original vulnerability has not changed.

\underline{\textbf{Need for patch quality assessment.}}
Let us assume a scenario where patch A does not pass PoV and patch B, which is repaired based on patch A, also fails to pass PoV. Which patch should be selected to be repaired in the next iteration? Existing iterative strategies choose patch B directly for the next iteration but cannot assess which patch is of better quality. In other words, they fail to compare which patch is more effective in stopping the vulnerability from being exploited.
This motivates us to assess the quality of all candidate patches. 
To quantify the effectiveness of patches in preventing the exploitation of vulnerabilities, we attempt to capture the dynamic information of each patch. Our main argument is that, \textit{under the same PoV, if exploiting patch A requires executing more statements compared with the original function B, it means that the repair attempt of patch A will affect PoV exploitation and continuously repair on patch A is more hopeful to fix the vulnerability}. Therefore, we select the widely used \textit{Statement Coverage}~\cite{DBLP:conf/icse/XiongLZ0018,KATCH,VarFix,DBLP:conf/icse/ChekamPTH17,DBLP:journals/jise/HwangLLL14} of taint statements to assess candidate patches.

\section{Approach}
\begin{figure}[!t]
    \centering
    \includegraphics[width=0.9\linewidth]{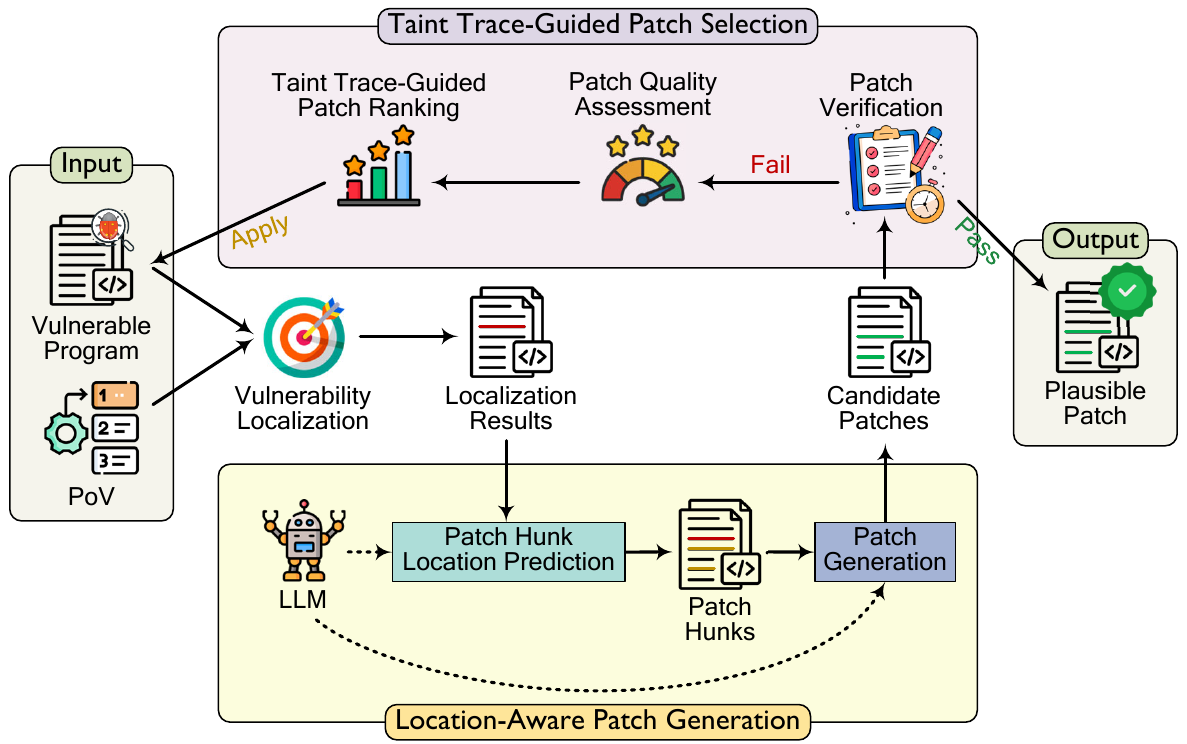}
    \caption{Workflow of \sysname.}
    \label{workflow}
\end{figure}

As shown in Figure~\ref{workflow}, \sysname includes four steps: \textit{Vulnerability Localization}, \textit{Location-Aware Patch Generation}, \textit{Taint Trace-Guided Patch Selection}, and \textit{Iterative Repair}. \sysname takes a vulnerable program with its PoV as input and outputs a plausible patch. 

\noindent \textbf{Step 1: Vulnerability Localization.}
Given the vulnerable program and its PoV, the localization tool first generates the localization results, which contain the vulnerable function and the vulnerable hunk locations.

\noindent \textbf{Step 2: Location-Aware Patch Generation.} 
 The \textit{patch hunk location prediction} component treats the location prediction task as a translation task and generates the sequence of line numbers that represent the patch hunks that need repair. Next, we construct a special prompt for each vulnerable function and utilize an LLM to generate several candidate patches.

\noindent \textbf{Step 3: Taint Trace-Guided Patch Selection.}  This step aims to verify candidate patches and assess the quality of failing-patches, ultimately outputting the top-ranked one. For each generated candidate patch, if it fails to pass PoV verification, we will assess its quality based on whether it introduces new vulnerabilities and its taint statement coverage. Then the \textit{taint trace-guided patch ranking} component ranks the assessed patches and selects the top-ranked patch for the subsequent iteration.

\noindent \textbf{Step 4: Iterative Repair.}  Given the top-ranked patch, we apply it to the vulnerable program. The iterative repair process contains \textit{Vulnerability Localization}, \textit{Location-Aware Patch Generation}, and \textit{Taint Trace-Guided Patch Selection}. The total process will be executed iteratively until one patch passes PoV verification or the iteration number reaches the user-defined $Num_{iteration}$.

\subsection{Vulnerability Localization}\label{3.0}
Given a vulnerable program and its PoV, we utilize a localization tool to generate the localization results, which contain the vulnerable function and the vulnerable hunk locations. The vulnerable function represents the function that has been detected with a vulnerability, while the vulnerable hunk locations refer to the locations of several consecutive statements that have been determined to be suspicious. Considering that fault localization is usually developed as an independent field and existing AVR techniques employ off-the-shelf fault localization tools in the repair pipeline, we do not discuss fault localization here and reuse CrashAnalysis~\cite{CrashRepair}.

\subsection{Location-Aware Patch Generation}\label{3.1} 

In this step, our goal is to automatically generate the candidate patches for each vulnerable function $F_{v}$. In order to effectively activate the performance of LLM, we utilize the Chain-of-Thought (CoT) method~\cite{DBLP:conf/emnlp/FanTLC0J23,DBLP:conf/coling/KumarMRKAD25,DBLP:conf/nips/Yuan0YZX0GWY24}, which provides information step by step. The information contains vulnerable hunk locations, predicted patch hunk locations, and few-shot examples. This step can be divided into three phases. 

\noindent \textbf{\underline{{Patch Hunk Location Prediction.}}}
According to our Observation 1 (cf. Section~\ref{motivation1}), it is important for LLMs to pay attention to the locations that need to be repaired before generating patches. Therefore, in this phase, we design a \textit{patch hunk location prediction} component that transforms the prediction task into a translation task. Specifically, the component takes as input both the vulnerable function and the identified vulnerable hunk locations. By processing this information, the component outputs a \textit{predicted patch hunk location sequence}. This sequence provides a detailed representation of the locations that need to be patched, thereby guiding the repair process more effectively. To achieve this, we employ an off-the-shelf LLM to act as a predictor. 

\begin{figure}
    \centering
    \includegraphics[width=\linewidth]{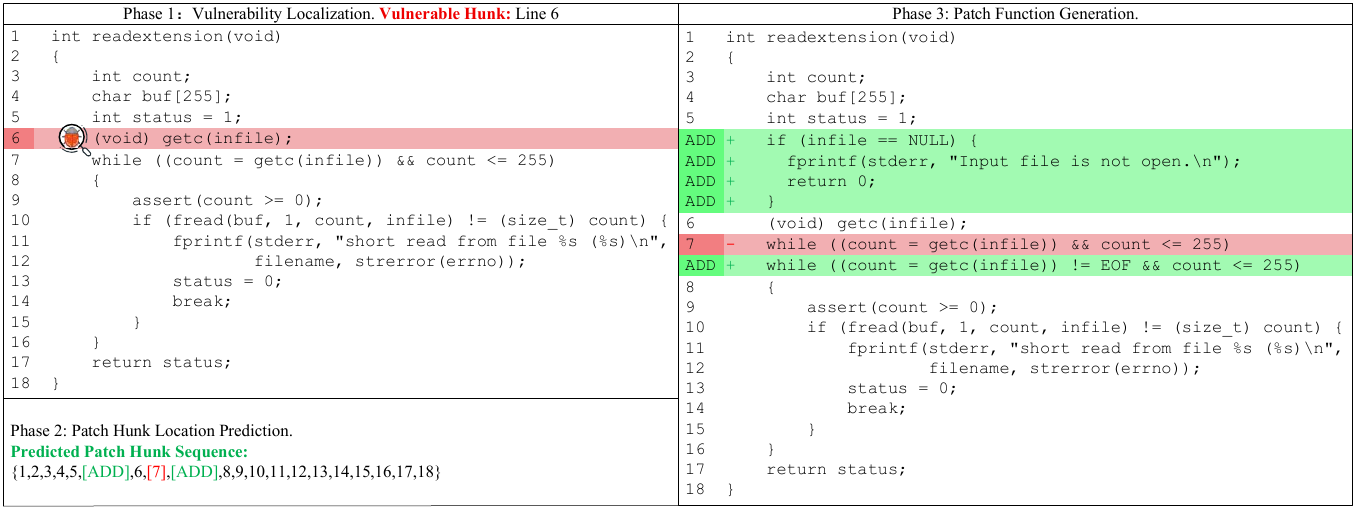}
    \caption{A running example (CVE-2016-3186~\cite{CVE-2016-3186}) of Location-Aware Patch Generation.}
    \label{RunningExample}
\end{figure}
Figure~\ref{RunningExample} shows a running example of CVE-2016-3186~\cite{CVE-2016-3186}, which illustrates how the \textit{patch hunk location prediction} component works. \underline{Line 6} in the \texttt{readextension} function is the vulnerable hunk provided by the localization tool, the \textit{patch hunk location prediction} component aims to predict the line number sequence, which represents that the locations need to be patched. Specifically, the output \textit{predicted patch hunk location sequence} of CVE-2016-3186 is:
\begin{equation}
    \{1,2,3,4,5,[ADD],6,[7],[ADD],8,9,10,11,12,13,14,15,16,17,18\} \nonumber
\end{equation}
where $[7]$ means to repair this vulnerability, we need to remove \underline{Line 7}. $[ADD]$ tokens indicate potential patch hunk locations within the function. These tokens specify locations where hunk needs to be removed/added to effectively repair the identified vulnerability. Other unchanged line numbers indicate that these lines do not need to be modified. It is important to note that the line numbers in \textit{predicted patch hunk location sequence} are aligned with the line numbers of the vulnerable function to ensure that there is no ambiguity during the patch generation process.

\begin{figure}
    \centering
    \includegraphics[width=1\linewidth]{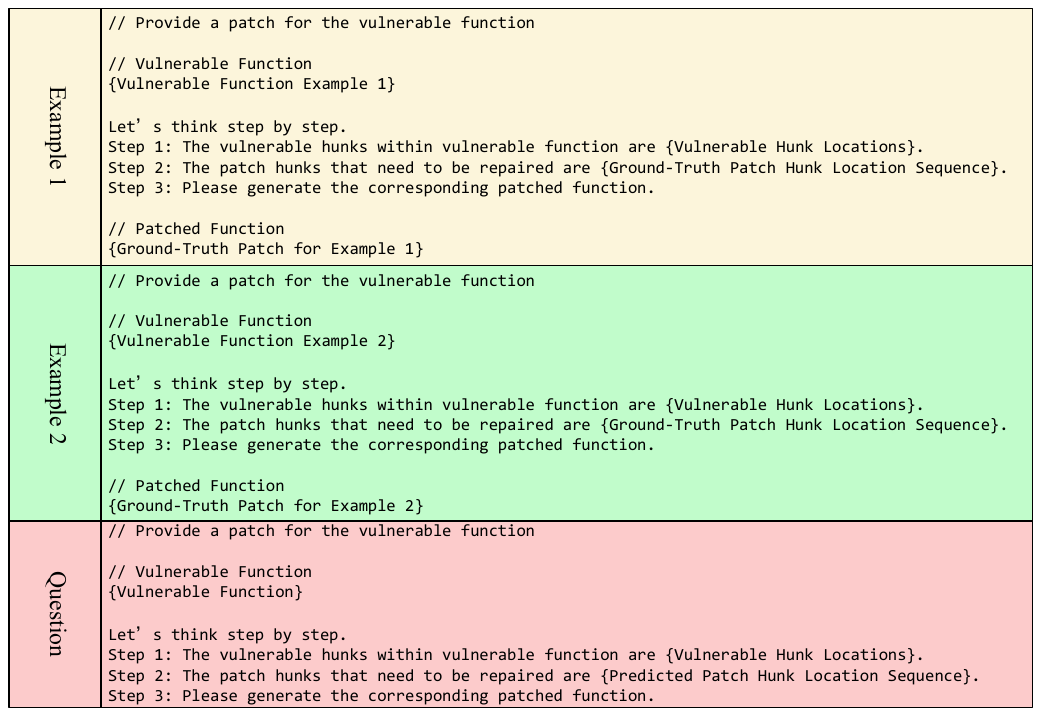}
    \caption{Chain-of-Thought prompt for each vulnerable function.}
    \label{CoT}
\end{figure}

\noindent \underline{\textbf{Prompt Construction.}} 
Previous works~\cite{ThinkRepair,DBLP:journals/corr/abs-2412-02906} have proven that few-shot learning can improve the performance of LLMs. Therefore, in this phase, we aim to collect few-shot examples and construct a special prompt for each vulnerable function $F_{v}$ to be repaired.
The prompt is constructed using a CoT template, which is illustrated in Figure~\ref{CoT}. It contains two parts: two few-shot examples that share the same CWE-ID with $F_{v}$ and the vulnerable function $F_{v}$. 
Our step-by-step reasoning chains consist of three steps. In Step 1, we provide the vulnerable hunk to help the LLM understand the cause of the vulnerability. In Step 2, we provide the \textit{patch hunk location sequence} to help the LLM identify which locations need to be patched. In Step 3, we allow the LLM to generate several patches to address the identified vulnerabilities.
The step-by-step reasoning chains for few-shot examples and vulnerable functions are constructed separately, as described below.
\begin{itemize}[leftmargin=1em]
    \item In order to generate step-by-step reasoning chains for two few-shot examples, we collect the <vulnerable function, patch function> pairs from the widely used Big-Vul~\cite{Big-Vul} dataset. We randomly select two <vulnerable function, patch function> pairs that share the same CWE-ID with $F_{v}$. The \{Vulnerable Hunk Locations\} and \{Ground-Truth Patch Hunk Location Sequence\} are obtained by executing \texttt{diff}-operation for each <vulnerable function, patch function> pair.
    
    \item For the vulnerable function $F_{v}$, the \{Vulnerable Hunk Locations\} is provided by the localization tool, and the \{Predicted Patch Hunk Location Sequence\} is generated in \textit{patch hunk location prediction} phase.
\end{itemize}

\noindent \underline{\textbf{Patch Function Generation.}} We utilize the LLM as a patch generator. To ensure that the model operates within the intended task scope, we initialize it with a system message explicitly defining its role: ``You are now playing the role of an automated vulnerability repair tool.''  The constructed CoT prompt is then used as user message to generate the patch function. A running example is provided in Figure~\ref{RunningExample}. 
The repair tool generates the patch hunk within the location of $[ADD]$ tokens and comments out the hunks (i.e., \underline{Line 7}, \underline{Lines 14-16}) that need to be removed.

\subsection{Taint Trace-Guided Patch Selection}\label{3.2}
In this step, we first verify whether generated candidate patches pass PoV verification. If all patches fail, we aim to assess the quality of each patch and select the top-ranked patch for the next iteration of repair. This step can be divided into three phases.

\noindent \textbf{\underline{{Patch Verification.}}}
\sysname first applies each candidate patch to the vulnerable program, then compiles and runs PoV to verify each candidate patch generated by LLM. Once the plausible patch that passes the PoV verification is found, the total repair process is terminated. 
If a candidate patch fails to pass all test cases, \sysname first collects the failing PoV information, which can aid LLM in understanding the failure causes and provide guidance to generate the correct fix. In particular, test failure messages can be divided into three categories: (1) Time Out, (2) Compile Fail, and (3) Failing PoV. 
For these failing patches, we immediately discard timeout candidate patches as they cannot provide any verification information. 
For patches with compilation errors and patches that fail PoV verification, we preserve them for subsequent ranking and analysis. Due to the fact that there are only 1-2 PoVs for each vulnerability, in the patch verification phase, we leverage the fuzzing part of \textsc{CrashRepair}~\cite{CrashRepair} to generate additional PoVs for mitigating test-overfitting.

\noindent \underline{\textbf{Patch Quality Assessment.}} 
This phase aims to assess the quality of failing patches. As described in Observation 2 (cf. Section~\ref{motivation2}), we have established the significance of patch quality assessment, especially under scenarios where none of the candidate patches in the current iteration successfully fixes the vulnerability. Our primary goal is to identify (1) whether the failing patches introduce new vulnerabilities and (2) the taint statement coverage of the failing patches. Therefore, we apply each failing patch to the vulnerable program and utilize CrashAnalysis~\cite{CrashRepair} to extract the taint trace and predict the CWE type. A patch is considered to not introduce new vulnerabilities if it meets two conditions: \ding{182} The CWE type of the vulnerability in the patch is the same as in the original program, which means there is no new vulnerability introduction. \ding{183} The taint sink of the patch matches the taint sink of the original program, which means that the original vulnerability has not changed. If a patch introduces a new vulnerability, it will be discarded immediately.
For the taint statement coverage ($TSC$), we calculate it for each failing patch, which is formally defined as:
\begin{equation}
    TSC = \frac{Number\;of\;Executed\;Statements\;in\;Taint\;Trace}{Total\;Number\;of\;Statements} \nonumber
\end{equation}

\noindent \underline{\textbf{Taint Trace-Guided Patch Ranking.}} The goal of this phase is to select the candidate patch that has progressed towards repairing the vulnerability without introducing new vulnerabilities. Since all patches that introduce new vulnerabilities have been removed, we directly sort the remaining failing patches. Specifically, we rank them in descending order according to $TSC$ of each patch. The top-ranked patch will be selected as the input for Iterative Repair.

\subsection{Iterative Repair}\label{3.3}
In this step, we aim to iteratively repair the patch that fails to pass PoV verification. Specifically, we first apply the top-ranked failing-patch to the original vulnerable program. Following this, \textit{Vulnerability Localization} is conducted to identify the vulnerable function and vulnerable hunk locations. After that, the \textit{Location-Aware Patch Generation} step generates several candidate patches. 
The \textit{Taint Trace-Guided Patch Selection} step verifies the generated patches first. If one patch passes all PoVs, the iterative repair process is terminated. If not, we will assess the quality of failing patches.
Due to the fact that with the increasing of the iteration number, the total number of failing patches will increase exponentially, which will significantly harm efficiency. In order to solve this issue, we propose to maintain a failing-patch pool to store all the PoV-failing patches and compilation-failing patches across all iterations. 
The top-ranked failing patch will be selected for the next iteration, and it will be removed from the failing-patch pool.
This entire process continues until the number of iterations reaches the user-defined $Num_{iteration}$.

\section{Experimental Design}
We evaluate \sysname on the following research questions:
\begin{itemize}[leftmargin=1em]
    \item \textbf{RQ1: How does the performance of \sysname compare against the state-of-the-art approaches in generating plausible patches?} We aim to answer whether and to what extent can \sysname outperform the state-of-the-art baselines in generating plausible patches.
    
    \item \textbf{RQ2: How does the performance of \sysname compare against the state-of-the-art approaches in generating correct patches?} We aim to answer whether and to what extent can \sysname outperform the state-of-the-art baselines in generating correct patches. 

    \item \textbf{RQ3: What are the contributions of different components of \sysname in improving repair effectiveness?} We perform two sets of ablation studies to understand how two components (i.e., \textit{patch hunk location prediction} and \textit{taint trace-guided patch ranking}) impact the effectiveness of \sysname.
    
\end{itemize}

\subsection{Datasets}
Our evaluation dataset VulnLoc+ consists of 41 vulnerabilities from the VulnLoc~\cite{VulnLoc} benchmark and 12 vulnerabilities collected by Shariffdeen et al.~\cite{CrashRepair}, which together include a diverse set of 53 vulnerabilities related to Buffer-Overflow (BO), Divide-by-Zero (DZ), Integer-Overflow (IO), Null-Pointer-Dereferences (NPD), Data-Type-Overflow (DTO), and Use-After-Free (UAF). Of the 53 vulnerabilities in our dataset, 13 vulnerabilities cannot have their localization information generated with the localization tool and therefore cannot undergo further repair operations. Therefore, we use the remaining 40 vulnerabilities in our evaluation. Among these vulnerabilities, only nine vulnerabilities are single-hunk vulnerabilities, while the remaining 31 are multi-hunk vulnerabilities. We opt not to use widely adopted datasets such as Big-Vul~\cite{Big-Vul} and CVEFixes~\cite{CVEFixes} because they do not provide available PoVs.

\begin{table*}[htbp]
\centering
\caption{The number of plausible patches and correct patches generated by \sysname and baselines. GPT-4o-mini is used as the backbone model for all LLM-based approaches. ``\plausible'' indicates a plausible but incorrect patch. ``\correct'' denotes a correct patch. ``a/b'' denotes that the approach totally generates ``b'' plausible patches, with ``a'' of them being correct.}
\resizebox{\textwidth}{!}{
\begin{tabular}{lclcccccccccccccc}
\toprule
\multicolumn{3}{c}{Category} & \multicolumn{2}{c}{Localization Results} & \multicolumn{4}{c}{NMT-based Baselines} & \multicolumn{2}{c}{PA-based Baselines} & \multicolumn{3}{c}{LLM-based Baselines} & \multicolumn{3}{c}{\cellcolor{customgray}Ours: \sysname} \\
\cmidrule(lr){1-3} \cmidrule(lr){4-5} \cmidrule(lr){6-9} \cmidrule(lr){10-11} \cmidrule(lr){12-14} \cmidrule(lr){15-17}

Project & Type & Vul-ID & Function-Level Right? & Predicted Location Right? & VRepair~\cite{VRepair} & VulRepair~\cite{VulRepair} & VQM~\cite{VQM} & VulMaster~\cite{VulMaster} & \textsc{CrashRepair}~\cite{CrashRepair} & \textsc{VulnFix}~\cite{VulnFix} & ThinkRepair~\cite{ThinkRepair} & ChatRepair~\cite{ChatRepair} & ITER~\cite{ITER} & $Num_{iteration}$=1 & $Num_{iteration}$=2 & $Num_{iteration}$=3 \\
\cmidrule(lr){1-3} \cmidrule(lr){4-5} \cmidrule(lr){6-9} \cmidrule(lr){10-11} \cmidrule(lr){12-14} \cmidrule(lr){15-17}
\multirow{3}[0]{*}{BinUtils} & BO    & CVE-2017-6965 & \plausible     & \plausible     &   &   &   &   &   &   &   &   &   &   &   &\\
     & IO    & CVE-2017-14745 & \plausible     & \plausible     &   &   &   &   &     & \plausible     &   &   &   &   &   &\\
     & DZ    & CVE-2017-15025 & \correct     & \plausible     &   &   &   &   & \plausible     & \correct     &   &   &   &   &   &\\
\cmidrule(lr){1-3} \cmidrule(lr){4-5} \cmidrule(lr){6-9} \cmidrule(lr){10-11} \cmidrule(lr){12-14} \cmidrule(lr){15-17}
\multirow{2}[0]{*}{CoreUtils} & BO    & gnubug-19784 & \plausible     & \plausible     &   &   & \correct     & \correct     & \correct     & \plausible     &\plausible    & \correct     &\plausible    &\plausible    &\plausible    &\plausible\\
     & BO    & gnubug-25023 & \plausible     & \plausible     &   &   &   &   & \plausible     & \plausible     &   &   &\plausible    &   &   &\\
\cmidrule(lr){1-3} \cmidrule(lr){4-5} \cmidrule(lr){6-9} \cmidrule(lr){10-11} \cmidrule(lr){12-14} \cmidrule(lr){15-17}
\multirow{4}[0]{*}{Jasper} & BO    & CVE-2016-8691 & \plausible     & \plausible     &   &   &   &   & \plausible     & \plausible     &\plausible    &\plausible    &\plausible    &\plausible    &\plausible    &\plausible\\
     & IO    & CVE-2016-9557 & \correct     & \correct     &   &   &   &   & \plausible     &     &\plausible    &   &   &\plausible    &\plausible    &\plausible\\
     & BO    & REDTEAM-CVE-2020-27828 & \plausible     & \plausible     &   &   &   &   & \plausible     &     &   &\plausible    &\plausible    &   &   &\\
     & BO    & REDTEAM-CVE-2021-3272 & \plausible     & \plausible     &   &   &   &   &   &   &     &     &   &   &   & \\
\cmidrule(lr){1-3} \cmidrule(lr){4-5} \cmidrule(lr){6-9} \cmidrule(lr){10-11} \cmidrule(lr){12-14} \cmidrule(lr){15-17}
LibArchive & IO    & CVE-2016-5844 & \correct     & \correct     &   &   &   &   & \plausible     &     &\plausible    &\plausible    &\plausible    &\plausible    &\plausible    &\plausible\\
\cmidrule(lr){1-3} \cmidrule(lr){4-5} \cmidrule(lr){6-9} \cmidrule(lr){10-11} \cmidrule(lr){12-14} \cmidrule(lr){15-17}
\multirow{2}[0]{*}{LibJPEG} & BO    & CVE-2012-2806 & \correct     & \correct     &\plausible    &\plausible    &   &\plausible    & \plausible     & \plausible     & \correct     &\plausible    &\plausible    & \correct     & \correct     & \correct \\
     & NPD   & CVE-2017-15232 & \correct     & \correct     &   &   &   &   & \plausible     & \correct     &   &   &   &   &   &\\
\cmidrule(lr){1-3} \cmidrule(lr){4-5} \cmidrule(lr){6-9} \cmidrule(lr){10-11} \cmidrule(lr){12-14} \cmidrule(lr){15-17}
\multirow{3}[0]{*}{LibMING} & BO    & CVE-2016-9264 & \correct     & \correct     &   &   &   &\plausible    & \plausible     & \plausible     & \correct     & \correct     &\plausible    &\plausible    &\plausible    &\plausible\\
     & UAF   & CVE-2018-8806 & \correct     & \correct     &   &   &   &   &   &   &     &     &   &\plausible    &\plausible    &\plausible\\
     & UAF   & CVE-2018-8964 & \plausible     & \plausible     &   &   &   &   &   &   &     &     &   &   &   &\\
\cmidrule(lr){1-3} \cmidrule(lr){4-5} \cmidrule(lr){6-9} \cmidrule(lr){10-11} \cmidrule(lr){12-14} \cmidrule(lr){15-17}
\multirow{16}[0]{*}{LibTIFF} & BO    & bugzilla-2633 & \correct     & \plausible     &   &   &   &   & \plausible     & \plausible     &\plausible    & \correct     &\plausible    & \correct     & \correct     & \correct \\
     & DZ    & bugzilla-2611 & \plausible     & \plausible     &   &   &   &   & \plausible     &     &\plausible    &   &   &\plausible    &\plausible    &\plausible\\
     & BO    & CVE-2016-10092 & \correct     & \correct     & \correct     & \correct     & \correct     & \correct     & \plausible     &     &\plausible    &\plausible    &   &   & \correct     & \correct \\
     & BO    & CVE-2016-10272 & \plausible     & \plausible     &   &   &   &   & \plausible     &     &   &   &   &   &   &\\
     & BO    & CVE-2016-3186 & \correct     & \correct     & \correct     & \correct     & \correct     & \correct     & \correct     &     & \correct     & \correct     & \correct     & \correct     & \correct     & \correct \\
     & BO    & CVE-2016-5314 & \plausible     & \plausible     &   &   &   &   &     &     &\plausible    &\plausible    &   &   &   &\\
     & IO    & CVE-2016-5321 & \plausible     & \plausible     &   &   &   &   & \plausible     & \plausible     &   &   &   &   &   &\plausible\\
     & BO    & CVE-2016-9532 & \correct     & \correct     &   &   &   &   & \plausible     &     &   &\plausible    &   &   & \correct     & \correct \\
     & BO    & CVE-2017-5225 & \plausible     & \plausible     &   &   &   &   & \plausible     & \plausible     &\plausible    & \correct     &\plausible    & \correct     & \correct     & \correct \\
     & DZ    & CVE-2017-7595 & \correct     & \correct     &   &   &   &   & \plausible     & \plausible     &   &   &   &   &   & \correct \\
     & IO    & CVE-2017-7601 & \correct     & \correct     &   &   &   &   & \plausible     & \plausible     &   &   &   & \correct     & \correct     & \correct \\
     & DTO   & CVE-2017-7599 & \correct     & \correct     &   &   &   &   &     &     & \correct     & \correct     & \correct     & \correct     & \correct     & \correct \\
     & DTO   & CVE-2017-7600 & \correct     & \correct     &   &   &   &   & \plausible     &     &   &\plausible    &   &\plausible    &\plausible    &\plausible\\
     & BO    & REDTEAM-CVE-2018-18557 & \correct     & \correct     &   &   &   &   &     &     &   &   &   &   & \correct     & \correct \\
     & BO    & REDTEAM-CVE-2022-4645 & \correct     & \correct     &   &   &   &   & \plausible     &     &\plausible    &   &   & \correct     & \correct     & \correct \\
     & BO    & REDTEAM-CVE-2022-48281 & \correct     & \correct     &\plausible    &\plausible    &\plausible    &\plausible    &     &     & \correct     &   & \correct     &   & \correct     & \correct \\
\cmidrule(lr){1-3} \cmidrule(lr){4-5} \cmidrule(lr){6-9} \cmidrule(lr){10-11} \cmidrule(lr){12-14} \cmidrule(lr){15-17}
\multirow{5}[0]{*}{LibXML2} & BO    & CVE-2012-5134 & \correct     & \plausible     &   &   &   &\plausible    & \correct     & \correct     &\plausible    &   &   &   &\plausible  &\plausible\\
     & BO    & CVE-2016-1838 & \correct     & \correct     &   &   &   &   & \plausible     & \plausible     &\plausible    &   &   & \correct & \correct  & \correct \\
     & BO    & CVE-2016-1839 & \plausible     & \plausible     &   &   &   &   & \plausible     & \plausible     &\plausible    &   &   &\plausible    &\plausible    &\plausible\\
     & NPD   & CVE-2017-5969 & \correct     & \correct     &   &   &   &   &     & \correct     &   &   &   &   & \correct & \correct \\
     & BO    & REDTEAM-CVE-2016-1833 & \correct     & \correct     &   &   &   &   & \correct     &     &   &   &   & \correct    & \correct     & \correct \\
\cmidrule(lr){1-3} \cmidrule(lr){4-5} \cmidrule(lr){6-9} \cmidrule(lr){10-11} \cmidrule(lr){12-14} \cmidrule(lr){15-17}
Potrace & BO    & CVE-2013-7437 & \plausible     & \plausible     &\plausible    &\plausible    &\plausible    &\plausible    & \correct     & \plausible     &\plausible&   &   &\plausible&\plausible &\plausible\\
\cmidrule(lr){1-3} \cmidrule(lr){4-5} \cmidrule(lr){6-9} \cmidrule(lr){10-11} \cmidrule(lr){12-14} \cmidrule(lr){15-17}
\multirow{3}[0]{*}{ZzipLib} & BO    & CVE-2017-5974 & \plausible     & \plausible     &   &   &   &   &     & \plausible     &   &   &   &   &   &\\
     & BO    & CVE-2017-5975 & \plausible     & \plausible     &   &   &   &   &     & \correct     &   &   &   &   &   &\\
     & BO    & CVE-2017-5976 & \plausible     & \plausible     &   &   &   &   &     &     &   &   &   &   &   &\\
\cmidrule(lr){1-3} \cmidrule(lr){4-5} \cmidrule(lr){6-9} \cmidrule(lr){10-11} \cmidrule(lr){12-14} \cmidrule(lr){15-17}
\multicolumn{2}{c}{Total} & \multicolumn{1}{l}{40} &   22/40    &   19/40   & 2/5   & 2/5   & 3/5   & 3/8   & 5/27  & 5/20  & 5/19  & 6/14  & 3/12  & 9/19  & 14/25 & 15/27 \\
\bottomrule
\end{tabular}}
\label{tab:effectiveness}
\end{table*}%

\subsection{Baselines and Evaluation Metrics}
\underline{\textbf{Studied Baselines.}} We compare \sysname with nine state-of-the-art (SOTA) approaches, including four NMT-based SOTAs, two Program Analysis (PA)-based SOTAs and three LLM-based SOTAs. 
(1) For NMT-based approaches,
VRepair~\cite{VRepair} pre-trains the repair model on a bug-fix corpus and fine-tunes it using pairs of <vulnerable function, repair function>.
VulRepair~\cite{VulRepair} utilizes Byte-Pair Encoding (BPE) and CodeT5~\cite{CodeT5} to embed the full vulnerable function with the CWE-ID to generate patches.
VulMaster~\cite{VulMaster} adopts the Fusion-in-Decoder (FiD) framework~\cite{FID} to combine vulnerable code, its AST sequence, and external information to produce a repair patch.
VQM~\cite{VQM} employs a localization component to identify tokens that require repair before generating the patch.
(2) For Program Analysis-based approaches, we select \textsc{VulnFix}~\cite{VulnFix} and \textsc{CrashRepair}~\cite{CrashRepair}.
\textsc{VulnFix} is the SOTA search-based approach~\cite{SoK}, which proposes a counter-example-guided inductive inference procedure that identifies invariants at potential fix locations by mutating program states. \textsc{CrashRepair} is the SOTA constraint-based approach~\cite{CrashRepair} which leverages concolic execution to infer desirable constraints at specific program locations and then searches for code mutations that satisfy these constraints.
(3) For LLM-based approaches,
ChatRepair~\cite{ChatRepair} repairs a bug with a conversational LLM, which inputs the buggy code and its test failure information.
ITER~\cite{ITER}, which directly applies an iterative repair strategy, is similar to our work. Specifically, for ITER, since they do not open-source their datasets and source code, we are unable to pre-train and fine-tune their patch generator. Therefore, we directly use their inference framework and employ \textit{GPT-4o-mini} as the patch generator.
ThinkRepair~\cite{ThinkRepair} employs a few-shot learning strategy, which automatically selects the few-shot examples using a cluster model and iteratively refines the generated patches.

\noindent \underline{\textbf{Evaluation Metrics.}} Following previous work~\cite{Recoder,RewardRepair,CoCoNuT,ExtractFix,DLFix}, we adopt two widely used metrics: number of plausible patches and number of correct patches. A plausible patch is a patch that passes all PoVs~\cite{KNOD,CURE,CrashRepair}.
Furthermore, we manually review and identify the correct patches based on the ground truth provided by the developer. A plausible patch is considered correct if it is semantically equivalent to the developer-provided patch~\cite{ThinkRepair,SequenceR,Recoder,SelfAPR}.

\subsection{Implementation}\label{implementation}
In the \textit{Vulnerability Localization} step, many localization tools~\cite{ExtractFix,DBLP:journals/usenix-login/HuangLTJ20,GenProg} can be used. In this work, we did not improve the localization tool and directly used the existing localization tool CrashAnalysis~\cite{CrashRepair} to generate taint traces and provide localization information at the statement/function level.
In the \textit{Patch Hunk Location Prediction} and \textit{Patch Function Generation} steps, we selected \textit{GPT-4o-mini}~\cite{GPT-4o-mini} as the backbone model and generated 5 candidate patches for each vulnerable function in each iteration. 
The entire iterative process lasted for three rounds (i.e., $Num_{iteration}=3$).
All of our experiments were conducted with an Intel Core i7-12700KF CPU (3.6GHz), 128GB RAM on Windows Subsystem for Linux 2 (WSL2, specifically Ubuntu 22.04).

\section{Experimental Results}
\subsection{RQ1: SOTA Comparison on Plausible Patches}
\underline{\textbf{Experimental Setting.}} (1) For NMT-based approaches, following previous work~\cite{VRepair,VulRepair,VQM,VulMaster}, we pre-trained these baselines using the bug-fixing corpus~\cite{VRepair,VulRepair}, and then fine-tuned them using the vulnerability-repair corpus~\cite{Big-Vul,CVEFixes,MegaVul}. Furthermore, we removed duplicate samples present in the VulnLoc+ dataset during fine-tuning. 
(2) For the PA-based approaches, we reproduced their source code and reported the experimental results.
(3) For LLM-based approaches, we reproduced the workflows of three baselines based on their papers. 
To ensure fairness, we set the maximum iteration number to 3 for all iterative baselines and used \textit{GPT-4o-mini} as the backbone model for these three baselines, with a detailed discussion of LLM selection in Section~\ref{LLMConfiguration}. For the LLM-based APR baselines, we chose vulnerability types and crash constraints as the feedback for domain adaptation.
(4) We studied the impact of the iteration number $Num_{iteration}$ on our \sysname. 
Following Ye et al.~\cite{ITER}, we evaluated three variants corresponding to $Num_{iteration}$=1, 2, and 3. We limited our selection to these values because the time required for iterative repair increases significantly with a higher number of iterations. Furthermore, we manually checked whether the function-level localization results and the predicted patch hunk locations were the same as those provided by the developers. 
\textbf{Localization information.} To maintain evaluation fairness, we provided NMT-based, PA-based, and LLM-based approaches with localization results rather than providing \textit{perfect localization information}~\cite{VulMaster}. 

\noindent \underline{\textbf{Results.}} Table~\ref{tab:effectiveness} illustrates the number of plausible patches generated by \sysname and baselines. The localization results are shown in the `Localization Results' column. The `Function-Level Right?' column indicates whether the function-level results provided by the localization tool are accurate, while the `Predicted Location Right?' column shows whether the predicted patch hunk locations match developer-provided patch. For function-level localization results, 22 out of 40 are correct. However, even when incorrect localization results are provided, plausible patches can still be generated.
\sysname can automatically fix a total of 27 vulnerabilities.
Compared with NMT-based baselines, \sysname demonstrates superior performance across all projects. 
\sysname demonstrates comparable or superior repair performance compared with PA-based approaches, generating the same number of plausible patches as \textsc{CrashRepair} and 7 more than \textsc{VulnFix}.
\sysname outperforms LLM-based approaches, by generating 8$\sim$15 additional plausible patches.
The last three columns show that our taint trace-guided iterative repair strategy boosts the effectiveness. \sysname increases the number of plausible patches from 19 ($Num_{iteration}$=1) to 27 ($Num_{iteration}$=3), representing an increase of 8 more patches compared with a single repair attempt.

\noindent \underline{\textbf{Analysis.}}
(1) All NMT-based approaches are less effective than LLM-based approaches.
This is because the cloze-style strategy cannot handle multi-hunk vulnerabilities, which is consistent with our Observation 1 (cf. Section~\ref{motivation1}). Furthermore, in the VulnLoc+ dataset, 77.5\% (i.e., 31/40) of the vulnerabilities are multi-hunk vulnerabilities, which results in very low performance for NMT-based approaches.
(2) Compared with these three iterative LLM-based baselines, \sysname can generate more plausible patches, indicating that our improved iterative strategy is effective. Furthermore, as $Num_{iteration}$ increases, the number of plausible patches also rises, suggesting that iterative repair strategy is effective for the AVR task.
(3) Compared with these two PA-based approaches, \sysname is capable of generating considerable or even more plausible patches, indicating that the LLM-based iterative strategy is effective in repairing real-world vulnerabilities.
(4) Learning-based approaches perform worse in BinUtils, LibXML2, and ZzipLib projects, this is due to the low localization accuracy of the localization tool in these projects and the localization results will directly impact the repair results. 
(5) For the repair results, we observe that \sysname repairs 16/25 Buffer-Overflow, 4/5 Integer-Overflow, 2/2 Data-Type-Overflow, 1/2 Null-Pointer-Dereference, 3/4 Divide-by-Zero, and 1/2 Use-After-Free vulnerabilities. This indicates that \sysname has the ability in solving different types of vulnerabilities. 

\begin{answerbox}{
\textbf{Answer to RQ1:} \sysname is able to generate considerable or more plausible patches (8$\sim$22) than NMT-based, PA-based, and LLM-based baselines, demonstrating its effectiveness.
}
\end{answerbox}

\begin{figure}[!t]
    \centering 
      \begin{subfigure}{0.48\linewidth}
        \centering   
        \includegraphics[width=1\textwidth]{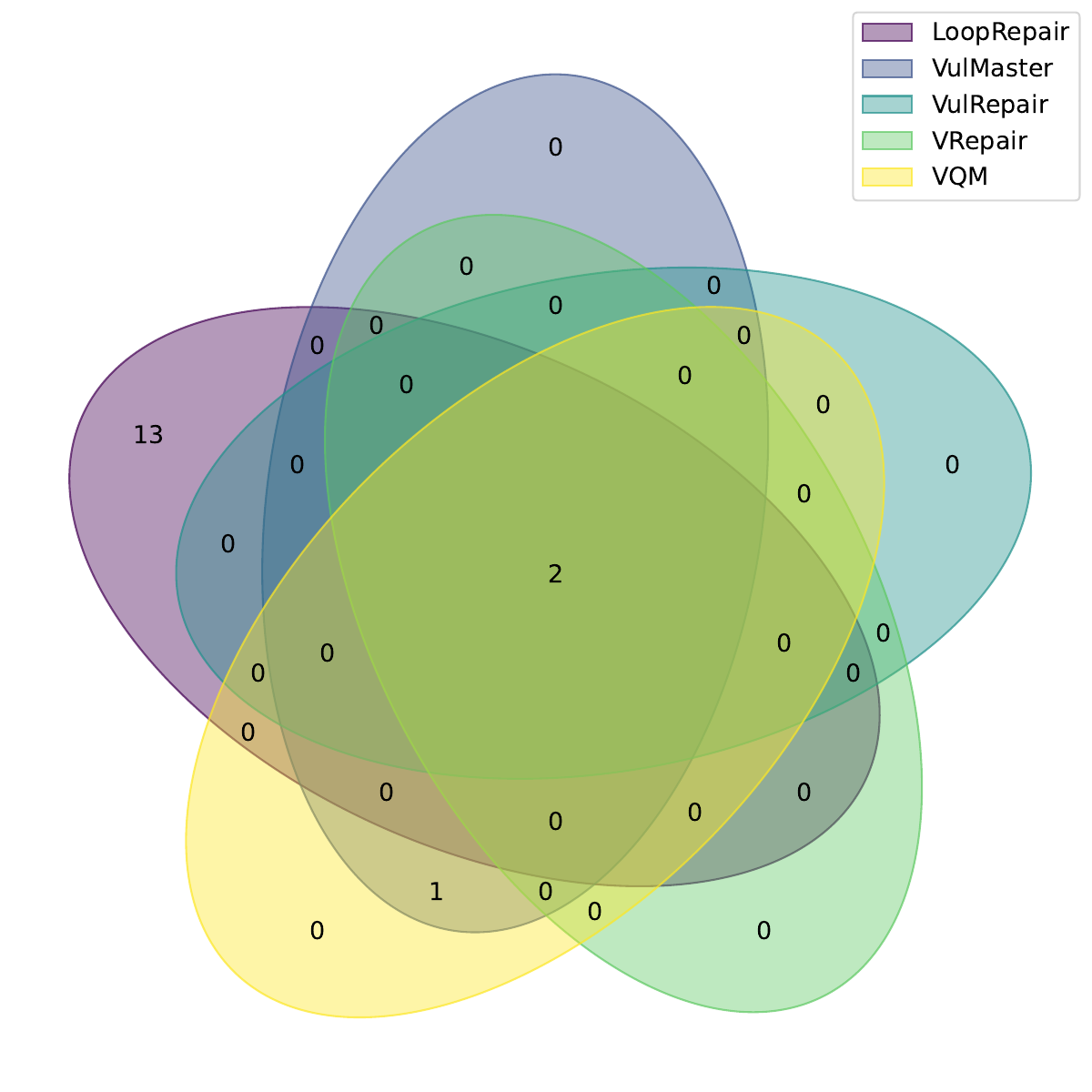}
          \caption{vs. NMT-based baselines.}
          \label{NMT_Venn}
      \end{subfigure} 
      \begin{subfigure}{0.48\linewidth}
        \centering   
        \includegraphics[width=1\textwidth]{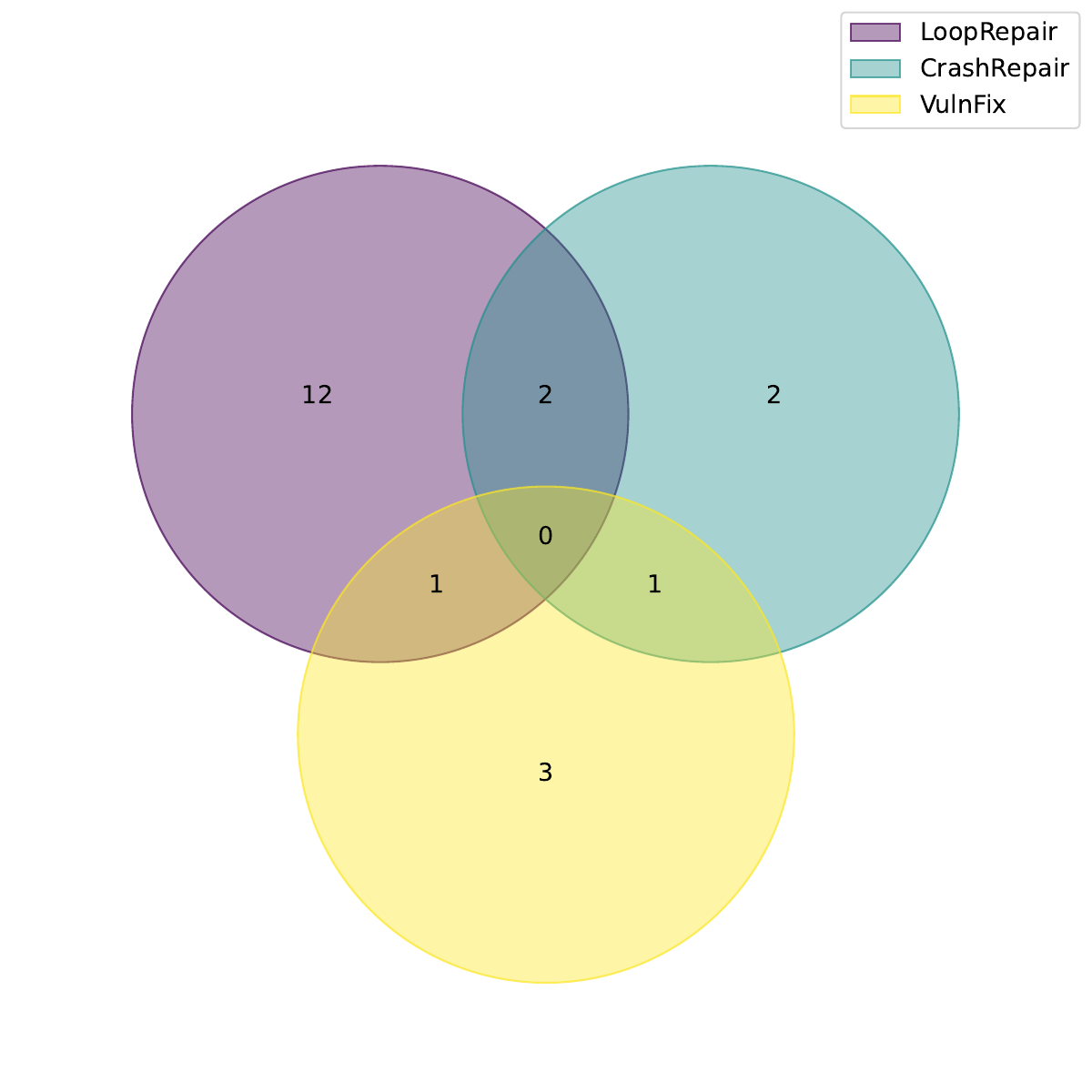}
          \caption{vs. PA-based baselines.}
          \label{PA_Venn}
      \end{subfigure} 
      \begin{subfigure}{0.48\linewidth}
        \centering   
        \includegraphics[width=1\textwidth]{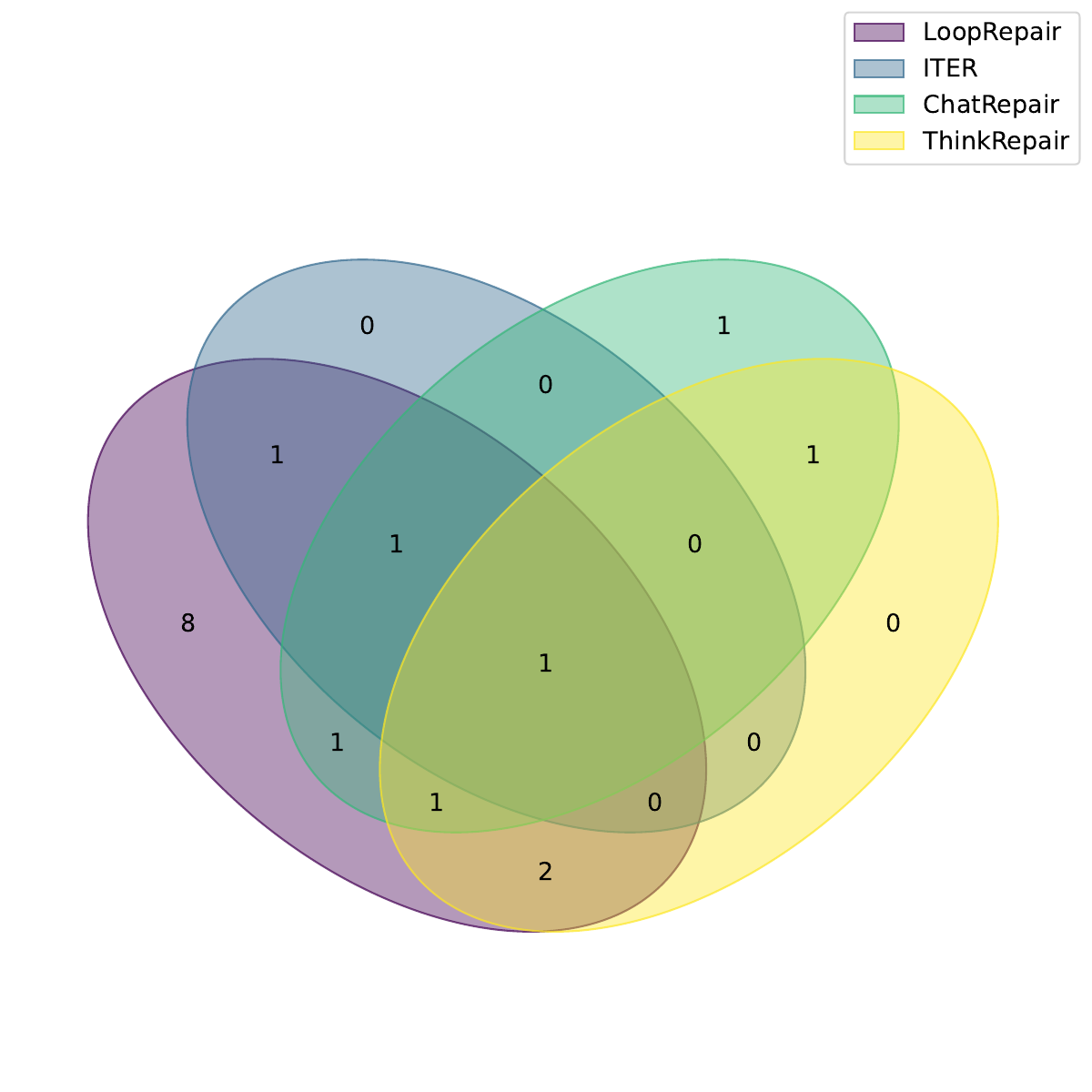}
          \caption{vs. LLM-based baselines.}
          \label{LLM_Venn}
      \end{subfigure}
  \caption{
  \label{VennGraph4NMT}
  Venn diagram of correct patch of \sysname vs. NMT-based, PA-based, and LLM-based baselines.
  }
\end{figure}

\subsection{RQ2: SOTA Comparison on Correct Patches}
\noindent \underline{\textbf{Experimental Setting.}} 
To further evaluate whether the generated plausible patch is correct, we invited \textit{n=3} volunteers (each with at least two years of experience in software engineering or software security) to assess the plausible patches generated in RQ1. Specifically, we first had two volunteers independently conduct patch correctness assessment and collect the results. For the patches that the two volunteers had different opinions, we invited a third volunteer to assess them. 

\noindent \underline{\textbf{Results and Analysis.}} Table~\ref{tab:effectiveness} shows the number of correct patches generated by \sysname and baselines, and we further drew three Venn diagrams to illustrate the performance difference between \sysname and baselines in vulnerability-repair. (1) Figure~\ref{NMT_Venn} presents the results of \sysname and four NMT-based baselines. It can be observed that \sysname achieves the best performance with a total of 15 correct patches, significantly outperforming the NMT-based baselines by generating 12$\sim$13 more correct patches (compared with VQM and VulMaster's 3 patches, and VRepair and VulRepair's 2 patches).
(2) Figure~\ref{PA_Venn} presents the results of \sysname and two PA-based baselines. Although the PA-based approaches can generate many plausible patches (20$\sim$27) through mutation strategies, most of these plausible patches (15$\sim$22) are incorrect. (3) Figure~\ref{LLM_Venn} shows the results of \sysname and LLM-based baselines. \sysname achieves the best performance, generating 9$\sim$12 more correct patches than the baselines (compared with ChatRepair's 6 and ITER's 3). Furthermore, the plausible patches generated during the iterative phase are more likely to be correct, with 6 out of 8 plausible patches being correct. This demonstrates that the iterative strategy effectively enhances the generation of correct patches. 
(4) For the vulnerabilities where the localization results are correct, we investigated these plausible but incorrect patches. We find that these patches are incorrect because the conditions in the added \texttt{if}-blocks are incomplete. For example, in the case of CVE-2017-7600 (Data-Type-Overflow), the generated patch only checks whether the pointer variable is \texttt{NULL}, but does not check if the value exceeds the upper limit of the \texttt{uint32} type, \texttt{0xFFFFFFFFU}. Although adding only the \texttt{NULL} check can fix the vulnerability under the specific PoV, we still consider this patch incorrect because it is not equivalent to the developer-provided patch.

\begin{answerbox}{
\textbf{Answer to RQ2:}
\sysname outperforms the baselines in generating correct patches, especially by repairing 8$\sim$13 unique vulnerabilities that other baselines fail to repair.}
\end{answerbox}

\subsection{RQ3: Ablation Study of \sysname}
\begin{table}[t]
  \centering
  \caption{Ablation study of \sysname.}
  \resizebox{0.47\textwidth}{!}{
    \begin{tabular}{lcr}
    \toprule
    Variants & \multicolumn{1}{c}{Plausible Patches} & \multicolumn{1}{c}{Correct Patches}\\
    \midrule
    \sysname & \textbf{27}  & \textbf{15}   \\
    \cmidrule(lr){1-3}
    -w/o patch hunk location prediction & 24 & 13\\
    -w/o taint trace-guided patch ranking & 19 & 7 \\ 
    -w/o both two components & 11 & 3\\
    \bottomrule
    \end{tabular}}
  \label{tab:ablation}
\end{table}

\noindent \underline{\textbf{Experimental Setting.}} 
\sysname has two important components: \textit{patch hunk location prediction} and \textit{taint trace-guided patch ranking}.
Therefore, in this RQ, we aim to conduct a comprehensive experiment to evaluate the impact of different components on \sysname’s performance. Specifically, we designed three variants of our \sysname. (1) \sysname without \textit{patch hunk location prediction}. (2) \sysname without \textit{taint trace-guided patch ranking}. (3) \sysname without both two components. All of our ablation experiments utilized the default settings outlined in Section~\ref{implementation}.

\noindent \underline{\textbf{Results and Analysis.}} The ablation results are shown in Table~\ref{tab:ablation}, we can observe that: (1) For \textit{patch hunk location prediction}, \sysname exhibits better vulnerability-repair performance compared with the variant without this component (i.e., 24→27 for plausible patches and 13→15 for correct patches). (2) The \textit{taint trace-guided patch ranking} component seems to contribute the performance of \sysname, which brings \textit{GPT-4o-mini} with a large improvement (i.e., 19→27 for plausible patches and 7→15 for correct patches). (3) Table~\ref{tab:effectiveness} shows that, among 40 vulnerabilities, 22 have correct function-level localization results. Of these 22, our approach correctly predicts the patch location for 19 vulnerabilities, resulting in 12 correct patches. In contrast, among the 21 vulnerabilities with incorrect patch locations, only 3 correct patches are generated. These results indicate that the \textit{patch hunk location prediction} component is helpful in generating correct patches.
(4) We further analyze the vulnerability types for which the repair fails after removal of each component. Removing \textit{patch hunk location prediction} component causes one Buffer-Overflow vulnerability and the unique repair for Data-Type-Overflow vulnerability to fail. We attribute this to the component accurately predicting all the locations that need to be patched, especially for types like Data-Type-Overflow where several locations need to be modified. Removing \textit{taint trace-guided patch ranking} component causes five Buffer-Overflow vulnerabilities, one unique Integer-Overflow, one unique Divide-by-Zero, and one unique Null-Pointer-Dereference vulnerability to fail. Successfully repairing these vulnerabilities requires the addition of strict judgment conditions to prevent the vulnerabilities from being triggered. We attribute this to the component's ability to accurately generate judgment conditions by selecting high quality patches for iterative repair.


\begin{answerbox}{
\textbf{Answer to RQ3:} The two components contribute substantially to \sysname, and their combination achieves the best performance. }
\end{answerbox}

\section{Discussion}

\subsection{Efficiency of \sysname}

\noindent \underline{\textbf{Experimental Setting.}} We evaluated the time consumption of \sysname on the VulnLoc+ dataset in generating 27 plausible patches and 13 failing patches generated in RQ1. The time consumption includes four steps: \textit{Vulnerability Localization}, \textit{Location-Aware Patch Generation}, \textit{Taint Trace-Guided Patch Selection}, and \textit{Iterative Repair}. 
\begin{figure}
    \centering
    \includegraphics[width=\linewidth]{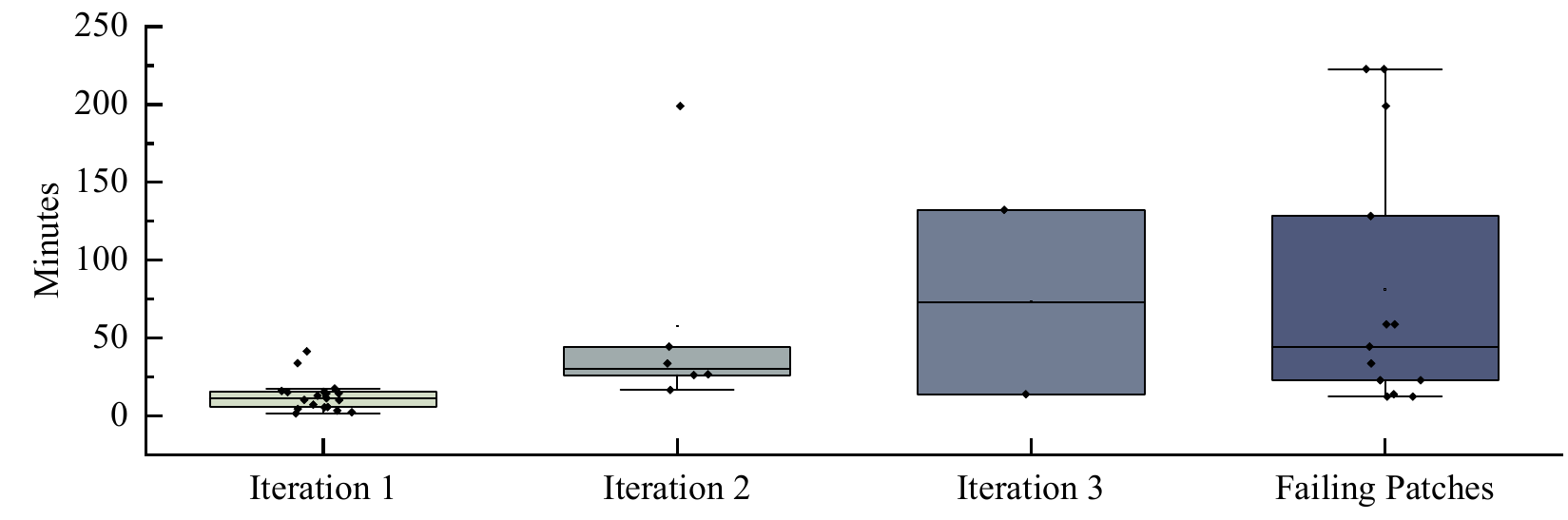}
    \caption{Time consumption of \sysname.}
    \label{fig:efficiency}
\end{figure}

\noindent \underline{\textbf{Results and Analysis.}} Figure~\ref{fig:efficiency} shows the time consumption distribution of 27 plausible patches and 13 failing patches discussed in RQ1. The x-axis indicates the number of inference iterations and the y-axis indicates the minutes spent on generating patches. The majority of plausible patches are generated in the range of 1-20 minutes. The time consumption increases with the number of inference iterations, affecting both the median value and the mean value. Failing patches consume a lot of time because they require running through three iterations.
To further analyze time consumption quantitatively, we recorded the minimum, median, and maximum values for plausible patches.
The time consumption ranges from 1.57 minutes to a maximum of 199.01 minutes. Half of the plausible patches are generated within 14.51 minutes. In addition, 77.78\% and 92.59\% of plausible patches are generated within 30 and 60 minutes, respectively. As reported by Noller et al.~\cite{DBLP:conf/icse/NollerS0R22}, a one-hour limit is considered realistic and tolerable for developers. It indicates that \sysname can efficiently repair real-world vulnerabilities.

\begin{answerbox}{
\textbf{Summary:}
Half of the plausible patches from \sysname are generated within 14.51 minutes. 92.59\% of plausible patches can be generated within one hour.}
\end{answerbox}

\subsection{Data Leakage Assessment}
\noindent \underline{\textbf{Experimental Setting.}} In order to evaluate the potential data leakage issue for large language models, we selected vulnerabilities from the continuously updated ARVO dataset~\cite{ARVO} that were fixed after the release of GPT-4o-mini-2024-07-18 (released on July 18, 2024). These vulnerabilities were fixed after the release of the LLM, so they are not part of the LLM training set. As a result, we obtained a total of 5 reproducible Java vulnerabilities. All experimental settings were the same as those in RQ1, and we used GPT-4o-mini-2024-07-18 as the backbone model for all baselines. We chose plausible patch number as our evaluation metric because none of these five vulnerabilities has published links to the developer-provided patches.

\noindent \underline{\textbf{Results and Analysis.}} The results of the data leakage assessment are presented in Table~\ref{tab:dataleakage}. (1) Based on the results, we conclude that \sysname demonstrates the ability to repair vulnerabilities in unseen programs, not just in those that might suffer from data leakage. Furthermore, \sysname outperforms all baselines, underscoring the practicality and viability of our approach. (2) Additionally, we manually evaluated the low-quality patches generated in the first iteration to investigate how much these low-quality patches impact the final results. For our \sysname, 13 out of 25 patches generated in the first iteration are low-quality. They either fail to compile due to obvious syntax errors (such as undeclared variables or functions), or introduce one or even more new vulnerabilities. 
The total iterative repair process will be affected if all the patches generated in the first iteration are low-quality.
Vulnerability \texttt{42537716} is particularly  affected because all patches generated in the first iteration are low-quality, which ultimately causes the repair process to fail. For the other four vulnerabilities, although our approach generates some low-quality patches (8 out of 20) in the first iteration, the presence of the patch selection component ensures that these patches do not affect the subsequent iterations.

\begin{answerbox}{
\textbf{Summary:} Data leakage does not significantly affect the performance of \sysname.}
\end{answerbox}

\begin{table}[!t]
  \centering
  \caption{The number of plausible patches generated by \sysname and baselines on unseen vulnerabilities.}
  \resizebox{0.48\textwidth}{!}{
    \begin{tabular}{lcccccc}
    \toprule
    \multicolumn{3}{c}{Vulnerability} & \multicolumn{3}{c}{LLM-based Baselines} & \multicolumn{1}{c}{\multirow{2}[2]{*}{\makecell{Ours: \\ \sysname}}}\\
    \cmidrule(lr){1-3} \cmidrule(lr){4-6} 
    Project & Language & \multicolumn{1}{c}{OSS Fuzz-ID} & ChatRepair~\cite{ChatRepair} & ThinkRepair~\cite{ThinkRepair} & ITER~\cite{ITER}  &\\
    \cmidrule(lr){1-3} \cmidrule(lr){4-6} \cmidrule(lr){7-7} 
    Apache-poi & Java & \multicolumn{1}{c}{390461322} &  &  &  & \correct \\
    \cmidrule(lr){1-3} \cmidrule(lr){4-6} \cmidrule(lr){7-7}
    Apache-poi & Java & \multicolumn{1}{c}{42537720} & & & & \correct \\
    \cmidrule(lr){1-3} \cmidrule(lr){4-6} \cmidrule(lr){7-7}
    Apache-poi & Java & \multicolumn{1}{c}{42537716} & & & &\\
    \cmidrule(lr){1-3} \cmidrule(lr){4-6} \cmidrule(lr){7-7}
    Apache-poi & Java & \multicolumn{1}{c}{42528888} & \correct     & \correct  & \correct     & \correct \\
    \cmidrule(lr){1-3} \cmidrule(lr){4-6} \cmidrule(lr){7-7}    
    Apache-poi & Java & \multicolumn{1}{c}{42528505} & & \correct  & \correct     & \correct \\
    \bottomrule
    \end{tabular}}
  \label{tab:dataleakage}
\end{table}%

\subsection{LLM Configuration Selection}\label{LLMConfiguration}
\noindent \underline{\textbf{Experimental Setting.}} To investigate the impact of different LLM configurations on our \sysname, we selected several alternative off-the-shelf LLMs, specifically \textit{GPT-3.5-turbo}~\cite{GPT-3.5}, \textit{Claude-3-5-sonnet}~\cite{Claude-3-5-sonnet}, \textit{Gemini-1.5-flash-002}~\cite{Gemini-1.5-flash-002}, and \textit{DeepSeek-V3}~\cite{DeepSeek-V3}. Each of these models was used to replace the \textit{GPT-4o-mini}~\cite{GPT-4o-mini} model utilized in our previous experiments. Notably, all other components and experimental settings were kept consistent across these different configurations to ensure a fair comparison of results. Additionally, we monitored the API costs in generating plausible patches for each model. This data provided insights into the cost-effectiveness of each LLM configuration.

\begin{table}
    \centering
    \caption{Total cost and effectiveness for different LLMs.}
    \resizebox{0.47\textwidth}{!}{
    \begin{tabular}{lccr}
        \toprule
        LLM Configuration & Plausible Patches & Cost & \makecell{Average Cost}\\
         \midrule
        GPT-4o-mini~\cite{GPT-4o-mini} & 27 & \$1.99  & \$0.074\\

        GPT-3.5-turbo~\cite{GPT-3.5} & 17 & \$10.04  & \$0.591\\

        Claude-3-5-Sonnet~\cite{Claude-3-5-sonnet} & 9 & \$48.87  & \$5.430 \\

        Gemini-1.5-flash-002~\cite{Gemini-1.5-flash-002} & 20 & \$1.214 & \$0.061\\

        DeepSeek-V3~\cite{DeepSeek-V3} & 22 & \$2.24  & \$0.102\\
        \bottomrule
    \end{tabular}} 
    \label{tab:cost_effectiveness}
\end{table}

\noindent \underline{\textbf{Results and Analysis.}} 
The results are presented in Table~\ref{tab:cost_effectiveness}. \textit{GPT-4o-mini} generates the highest number of plausible patches (i.e., 27) at a low total cost of \$1.99, resulting in a cost of \$0.074 per patch. In contrast, \textit{Claude-3-5-sonnet} produces only 9 patches at a high cost of \$48.87, leading to a cost of \$5.430 per patch. 
We analyzed the types of vulnerabilities that \textit{Claude-3-5-Sonnet} fails to repair and the results show that it fails to repair all 4 Divide-by-Zero vulnerabilities. This is a less frequent vulnerability type, and we suspect that this may be because there is less divide-by-zero-related data in the model's training set.
The remaining three LLMs, \textit{GPT-3.5-turbo}, \textit{Gemini-1.5-flash-002}, and \textit{DeepSeek-V3}, exhibit comparable performance and cost, ranging from 17 to 22 plausible patches generated, with costs from \$0.061 to \$0.591 per patch.
These results highlight the importance of the LLM configuration for effective patch generation, leading us to select \textit{GPT-4o-mini} as our backbone model due to its high performance and relatively low cost.

\begin{answerbox}{
\textbf{Summary:} \textit{GPT-4o-mini} is the most cost-effective option, generating the most plausible patches for a relatively low cost. In this work, we chose \textit{GPT-4o-mini} as our backbone model.}
\end{answerbox}

\subsection{Fitness Function Selection}\label{FitnessFunction}
\noindent \underline{\textbf{Experimental Setting.}} In the \textit{taint trace-guided patch selection} component, we utilize \textit{taint statement coverage} as the fitness function. To investigate the impact of different fitness function configurations on our \sysname, we selected three alternative fitness functions, specifically \textit{branch coverage}, \textit{AST difference}, and \textit{cosine similarity}. They are all widely used methods for assessing patches based on static code features~\cite{DBLP:conf/kbse/WangWLWQZMJ20,KATCH,DBLP:conf/kbse/XinR17}. Each of these fitness functions was used to replace the \textit{taint statement coverage} utilized in RQ1. Notably, all other components and experimental settings were kept consistent across these different configurations to ensure a fair comparison of the results.

\noindent \underline{\textbf{Results and Analysis.}} Table~\ref{tab:fitnessfunction} shows the performance of different fitness functions. \textit{Taint statement coverage} can generate the highest number of plausible patches (27) and correct patches (15). For the other three fitness functions, they show similar performance (17$\sim$19 plausible patches, 7$\sim$8 correct patches). Furthermore, we counted the number of patches generated in the first and subsequent iterations for different fitness functions. The results show that only \textit{taint statement coverage} is able to generate patches in subsequent iterations and the plausible patches generated in subsequent iterations are more likely to be correct, while others fail. In order to further investigate why only \textit{taint statement coverage} is effective, we analyzed the repair patterns of all 27 generated plausible patches. We found that 23 out of 27 plausible patches prevent vulnerabilities from being triggered by adding an \texttt{if}-block, while the remaining four are repaired by adding checks for the \texttt{if}-condition or \texttt{loop}-condition. Adding an \texttt{if}-block increases the statement coverage of taint traces, allowing \textit{taint statement coverage} to  capture this type of repair pattern effectively. Branch coverage cannot capture this type of pattern because the failed patches with an added \texttt{if}-block do not execute the \texttt{if}-branches. AST difference and cosine similarity also cannot capture the repair pattern with an added \texttt{if}-block.

\begin{table}[!t]
  \centering
  \caption{Effectiveness for different fitness functions.}
  \resizebox{0.47\textwidth}{!}{
    \begin{tabular}{lccccccr}
    \toprule
    \multirow{2}[5]{*}{Fitness Function} & \multicolumn{3}{c}{Plausible Patches} & \multicolumn{3}{c}{Correct Patches} \\
    \cmidrule(lr){2-4} \cmidrule(lr){5-7}
    & \makecell{First\\Iteration} & \makecell{Subsequent\\Iterations} & Total &  \makecell{First\\Iteration} & \makecell{Subsequent\\Iterations} & Total\\
    \cmidrule(lr){1-1} \cmidrule(lr){2-4} \cmidrule(lr){5-7} 
    taint statement coverage & 19 & 8 & 27 & 9 & 6 & 15  \\
    branch coverage & 17 & 0 & 17 & 7 & 0 & 7 \\
    AST difference & 19 & 0 & 19 & 8 & 0 & 8 \\
    cosine similarity &  18 & 0 & 18 & 7 & 0 &7 \\
    \bottomrule
    \end{tabular}}
  \label{tab:fitnessfunction}
\end{table}

\begin{answerbox}{
\textbf{Summary:} \textit{Taint statement coverage} is the most effective option, and we chose it as our fitness function.}
\end{answerbox}

\section{Related Work}
\noindent \textbf{\underline{Automated Vulnerability Repair.}}
Existing learning-based vulnerability repair approaches can be broadly categorized from three perspectives: NMT-based approaches, PA-based approaches and LLM-based approaches.

\noindent \textbf{NMT-based Approaches.} Learning-based approaches learn repair patterns from <vulnerable function, patch function> pairs~\cite{DBLP:conf/iwpc/ZhangXKY022,VRepair,VulRepair,VulMaster,VQM}. 
For example, Fu et al.~\cite{VulRepair} presented an automated software vulnerability repair approach, VulRepair, based on T5, which employed pre-training and BPE components to address some of the limitations of VRepair~\cite{VRepair}. 
Fu et al.~\cite{VQM} introduced a vision transformer-inspired vulnerability repair framework, VQM, which pinpointed vulnerable code regions by incorporating and leveraging Vulnerability Queries (VQs) before suggesting repairs. 

\noindent \textbf{PA-based Approaches.} Researchers have been dedicated to exploring vulnerability repair approaches based on program analysis~\cite{MemFix,DBLP:journals/usenix-login/HuangLTJ20,ExtractFix,SAVER,EffFix,DBLP:conf/pldi/ShariffdeenNGR21}. For example, ExtractFix~\cite{ExtractFix} treated the extracted vulnerability constraints as obligations that the synthesized patch had to satisfy and propagated these constraints to locations considered suitable for fixing. 

\noindent \textbf{LLM-based Approaches.} With the rise and development of large language models, researchers have begun a series of empirical studies to evaluate the performance of LLMs on real vulnerability repair tasks~\cite{DBLP:conf/sp/PearceTAKD23,DBLP:journals/corr/abs-2404-02525}.
Furthermore, several LLM-based Automated Program Repair (APR) approaches have demonstrated promising effectiveness in the Automated Vulnerability Repair (AVR) domain~\cite{AlphaRepair,ChatRepair,ITER,ThinkRepair}. For example, AlphaRepair~\cite{AlphaRepair} is an LLM-based repair tool that employs the CodeBERT~\cite{CodeBERT} with cloze-style APR, which means that it does not require finetuning on bug-fixing data. 

\noindent \textbf{\underline{SAST-Enhanced Learning-Based Approaches.}}
Researchers propose to combine static application security testing (SAST) with learning-based methods~\cite{VulMaster,VulDeePecker,MVD,DeepWukong,DBLP:journals/access/HegedusF22}.
For example, Zhou et al.~\cite{VulMaster} proposed VulMaster, which used the Fid framework to manage long vulnerable code and improve vulnerability repair by providing structural information from the AST.

\section{Threats to Validity}
\noindent \underline{\textbf{Internal Validity.}}
First, due to the limitations of the VulnLoc+ dataset, which provides only one or two PoVs for each vulnerability, we cannot use widely used spectrum-based localization tools~\cite{DBLP:conf/icse/KucukHP21,DBLP:journals/tse/WongGLAW16,DBLP:conf/kbse/JiangWXCZ19,DBLP:conf/kbse/AbreuZG09,DBLP:conf/qrs/LiuMZ20} for vulnerability localization. Additionally, we directly abandoned the vulnerabilities that fail to generate localization results.
Second, in order to maintain fairness, we choose the same LLM configuration for all LLM-based approaches~\cite{ITER,ChatRepair,ThinkRepair} in RQ1, which may lead to different results compared with their original approaches.
Third, when determining whether new vulnerabilities are introduced, we utilize CWE types generated by CrashAnalysis~\cite{CrashRepair}. Misclassifications of these CWE types can affect the effectiveness of our \sysname. In this work, we focus only on six significantly different vulnerability types, which reduce the likelihood of misclassification.
Fourth, if the patches generated in the first iteration are all low-quality, the performance of \sysname will be impacted. In our approach, we generated 5 patches in the first iteration to mitigate this impact.
Fifth, in RQ2, the manual evaluation was performed by three volunteers who have at least 2 years of experience. However, their expertise may not be sufficient for handling C/C++ programs, so we open-source these generated patches to address this issue.

\noindent \underline{\textbf{External Validity.}} 
To reduce the risk of an unrepresentative evaluation, we evaluate \sysname on the established VulnLoc dataset~\cite{VulnLoc} and 12 vulnerabilities collected by Shariffdeen et al.~\cite{CrashRepair}, which are called VulnLoc+, including six diverse types of vulnerabilities.

\section{Conclusion}
We propose \sysname, a novel automated vulnerability repair approach. \sysname provides the locations that need to be repaired to help LLMs focus on the patch hunk locations. Furthermore, \sysname improves the iterative strategy by incorporating taint trace-guided candidate patch ranking, which assesses the quality of generated patches and selects the top-ranked patch for the next repair iteration.  Our experimental results show that \sysname generates 27 plausible patches (comparable to or even 8 to 22 more plausible patches than the baselines), of which 15 are correct patches. In the future, we will seek to broaden the scope of the iterative repair process to increase the variability among the iterative branches.

\begin{acks}
This research was supported by the National Natural Science Foundation of China (No.62572421 and No.62202414), Postgraduate Research \& Practice Innovation Program of Jiangsu Province (No. KYCX25\_3968), the Open Foundation of Yunnan Key Laboratory of Software Engineering (No.2023SE201), and the Open Project of Industry-Education Integration Innovation Center of Specialized Cybersecurity (Yangzhou University) under Grant YZUCSC2025KF02.
\end{acks}

\clearpage
\bibliographystyle{ACM-Reference-Format}
\bibliography{sample-base}


\end{document}